# QUALITY INCREASES AS THE ERROR RATE DECREASES

Fabrizio d'Amore

Department of Computer, Control and Management Engineering
Sapienza University of Rome, Italy
damore@diag.uniroma1.it

## ABSTRACT

*In this paper we propose an approach to the design of processes and software that aims at decreasing human and software errors, that so frequently happen, making affected people using and wasting a lot of time for the need of fixing the errors. We base our statements on the natural relationship between quality and error rate, increasing the latter as the error rate decreases. We try to classify errors into several types and address techniques to reduce the likelihood of making mistakes, depending on the type of error.*

*We focus on this approach related to organization, management and software design that will allow to be more effective and efficient in this period where mankind has been affected by a severe pandemic and where we need to be more efficient and effective in all processes, aiming at an industrial renaissance which we know to be not too far and easily reachable once the path to follow has been characterized, also in the light of the experience.*

## 1. INTRODUCTION

This paper is not aiming at presenting an innovative computer science's result, rather it wants to make people more sensitive and ready to an approach to organization, governance and design aimed at greater effectiveness and efficiency, as deriving from the increase in the overall quality of the process under consideration, because of the reduction in the quantity of errors, of any nature. The main question we want to answer to is "how can digitalization best help innovation and development?" This isn't exactly research on computer science, rather is research on how computer science, or (better) information technology, should be interfaced to other human activities aiming at innovation and at supporting their development, and trying the use resources at best, and not for fixing errors or solving computer problems. The tools are old at least two decades but we have seen that, because of a strong motivation like the pandemic, solutions technically available twenty years ago have today allowed to increment our efficiency, being however yet far from the best we can obtain.

It is well-known that unexpected (often unintentional) events, can seriously damage any process, and the time spent recovering from the error is far greater than what it would have taken if the process would have gone without surprises. Further, ore we observe that way we recover from the error is frequently improvised due to a typical college education that does not focus on error handling.

Yet we consider important the subject and the recovery process, because an education at such subject, carried out using consolidated and tested effective methodologies, would lead to a better handling of the error situation. Hence the necessity to collect best practices and conceptual frameworks for obtaining a powerful set of instruments for error mitigation.

In this paper we'll often use the terms "user" and "operator." The two words should be meant similar, but we want to address the fact that the former is more generic, and referred to someone not expert of information technology, the latter is referred to a user that must accomplish simple task such as data entry. Also, we'll use the terms "computer science" and "information technology." The former is to describe a wide discipline, also having theoretical problems, and on which much research has being carried out; the latter references to some of the practical repercussions obtained from the former, which impact on human activities.

We wish to point out that we are not aware of any other work like this one, so we must present our model as new research that cannot cite previous literature nor make comparisons. Nevertheless, we feel that new research, addressing well-known problems of the transfer of computer science to information technology, deserves anyway to be considered, and looks extremely modern. In the same way, we cannot yet carry out experimental evaluations of what we propose, because what we are introducing is not yet so well defined as to be able to identify precise tools, although some can be easily imagined. However, part of our statements is referring to well-known issues and close to the common experience of many actors.

The paper has the following structure. Section 2 introduces an initial classification of errors and emphasizes the importance of digital hygiene; Section 3 presents a discussion on the topics introduced; and Section 4 draws some conclusions.

## 2. TYPES OF ERROR

Here we give a first classification, without claiming to be all-encompassing. The attempt is to take into consideration the most frequent or significant causes of error. We dedicate a subsection to each recognized cause.

### 2.1. Governance Errors

In many organizations, especially SMEs, the management is not educated at the information security, leaving decisions to be taken by IT people, often a handful of people. Yet, IT people are not educated at governance decisions, such choosing the appropriate policy for some given class of documents and are not completely aware of impact on the organization of the requirements of the information security [1]. For instance, choosing what category of documents should be confidential is a strategic decision and should not be taken by a technician; it is in fact the responsibility of the top management. Technicians will choose how to ensure confidentiality of some documents, like determining [2] the best level which to introduce the encryption in.

Another governance error is associating wrong or ambiguous requirements to information. This is a strategic error, and every organization should define a method for leaving users to let the management know such errors.

### 2.2 Operator Errors

Operators can apply policies in a wrong way. It is a mistake done during a procedure and (based on the professionalism of the operator) the organization should not define a standard countermeasure.

We believe that typos are a very common type of error; every human can make them. Hence, the need to minimize the writing at operator side, by letting the counter-interested, or the owner of the information, type it (probability of typos is much minor), and then help the data input by means of automated procedures. For instance, (s)he could type and check information, have an automatic tool producing a corresponding QR-code, to be shown to the operator, that should only use an automatic reader for inputting data. To this purpose it is appropriate to mention the huge research done on QR-codes and other types of bidimensional barcodes in the last decade, their improvement and efficientization; see e.g., [3, 4, 5, 6, 7, 8]. In general, for operators, the less they write, the better.

### 2.3. Omissions

To omit information or suitable details should be considered an error: often it is very expensive to retrieve proper information when one realizes its lack. The repetition of similar problems enables the operator to be able to predict in advance that it will need certain details, so that a retrieval procedure can be provided for time, perhaps by asking it to IT staff. The goal is to abolish every omission.

### 2.4. Software Issues

We know that various problems can arise from using applications: bugs, functions missing or hidden, etc. Of course, we don't want to tell software engineers how to design and test the software and its user-interface. However, we focus the fact that engineers view could be different from users/operators view. Therefore, we recommend that for the entire cycle of life of the software product some expert user/operator should complete the development team.

A typical source of issues is the duplicate input of information, or the input of information already available, perhaps on other software platform. All computer scientists know that this is increasing the probability of errors, due to possible problems of consistency and maintenance. With respect to this question, we propose an enlargement of the security by-design paradigm [9] by letting the design process include the awareness about other applications/platforms and we call it "awareness by-design." The same arguments that motivate the security by-design approach are at the base of the awareness by-design. This means that it should be a rule to design new software without ignoring pre-existing one and letting new and old platforms able to exchange data and avoiding any duplication. Of course, this is not always possible, especially due to the vendor lock-in policies, that prevent the open approach. To this matter we observe that the open approach should be pursued at any cost, not only for avoiding the vendor lock-in, but to make it simpler the exchange of data between platforms. Vendors will do not love this approach, because apparently in contrast with their profit goals, but this is a myopic vision because in the long run they will benefit from the open approach, the preferred one (or so it should be) by industries, public administration, universities, and all other organizations. Closed platforms should be abandoned, because strongly anti-economic.

Yet about software we mention the need of satisfying the information security requirements, and this can be done in several ways. However, we'd choose the cryptographic mode, especially if information-theoretically secure, rather a traditional approach at an application level, because an attacker can more easily make an application crashing, for some unexpected input or other application-level detail but cannot break solutions that are information-theoretically secure.

### 2.5. Inventory not Up to Date

In many cases information about hardware and software are incomplete. This is due to a fast-evolving situation, or to a provisional setting (only temporary), or other causes. A partial description of the inventory is the source of issues when handling critic events, like incident handling, update/upgrade handling, decisions about cloud services and others. At critical moments incomplete/wrong inventories could lead to neglect special or relevant cases, what could greatly increase the inefficiency and the waste of resources.

### 2.6. Incidents not Well-Managed

Incidents and anomalies should be managed keeping into account the appropriate information security requirements. Accounting and non-repudiation should be guaranteed in such a way that every action can be attributed to one subject, and this cannot repudiate it. Information on incidents/anomalies should be quickly collected and made available. Of course, an outdated inventory would make the handling much more complicated.

## 2.7. Lack Of Digital Hygiene

By "digital hygiene" we mean that part of cybersecurity that intersects the daily life of operators and users. Provocatively, we claim that a correct digital hygiene would make useless an antivirus, because some natural caution would do the job. And this hygiene should be spread as much, by addressing which behaviors could be virtuous and which could be reprehensible. We need real awareness, especially regarding the well-known social engineering, and passwords management. The view of complicated rules for creating a new password (e.g., a smaller-case letter, one capitalized, a figure, a special character, etc.) is at this point obsoleted: the increase in security with stronger passwords is completely (and beyond) balanced with the increase of insecure behaviors. Better to resort to alternative authentication methods, also based on multi-factor authentication, which have existed for some time but are struggling to overcome the traditional user/password scheme. Very instructive the Schneier's intervention [10, 11]. Of course, any education on digital hygiene should be mandatory, addressing caution behaviors and explaining social engineering (and not neglecting what phishing, spoofing, and ransomware are, etc.) and done inside the working time (in order not to incur a bad propensity). We should understand that what is obvious, if not trivial, for a computer scientist or a technologist is totally alien to a final user/operator.

Finally, we want to underline that confidentiality and authentication/integrity are very often requirements of the e-mail, but messages are too often left completely unprotected while stored in some server. An organization could easily implement the OpenPGP protocol [12], at zero cost, while managing only the public keys of local users in a centralized quasi-static manner (e.g., an LDAP) in such a way that all public keys of local users can be trusted, so to have an easy-to-use method of user encryption/signature, at least at a local level. Users should practice, when required, the user encryption, as discussed in [2].

## 3. DISCUSSION

In this paper two themes strongly interconnected are presented. First, reducing errors or issues, let us gain the real advantage of a full digitalization. Up to date, we haven't been careful enough, and in addition to errors, that are a component of human beings, we had to spend a lot of time (inefficiency!) in issues coming from the use of computers. Who didn't expend long time in recovering a password, just because it was forgotten and, for some reason, we weren't using a password manager? The traditional rules for creating a new password are too restrictive and we think that the cost of the total time wasted in password recovering is greater than the benefits coming from having a more secure password (that – don't forget – additionally pushes people towards incautious behaviors). In a broader sense, we need to eliminate any inefficiency or issue coming from digitalization. As a further example we saw the pandemic has forced us to make greater use of digital resources, which have already existed for several years. No new technologies but solutions that can be traced back to the last millennium. A gap between research and full technological transfer has always existed, but this isn't a good reason to accept it and make it a rule. Once more, we don't need issues coining from digitalization itself.

As for errors, in addition to adopting mechanisms to reduce them, we should stop dealing with them in an artisan and improvised way. Errors have existed for some long time, so we should predispose, from the very first education, methods that have proved successful in their resolution, as well as other best practices that help to reduce them. Who should take care of this? In our opinion, computer scientists, that are sensitive and qualified, have a mathematical method for designing and testing. We aren't saying that they are the only people prepared enough, but what it should be clear is that we need a multidisciplinary team, that offers a new type of specialization: the horizontal extent is the new vertical depth.

The second theme, strongly interconnected with digitalization, is the digital hygiene. Yes, this is a part of cybersecurity, but all users should be able to understand it. The modern meaning of cybersecurity is "computer security" (this is how Wikipedia re-directs the word "cybersecurity") and a correct hygiene is at its basis. This hygiene should extensively cover subjects like password management, social engineering, inserting USB devices, phishing and spoofing, net-etiquette, privacy, ransomware, etc. Certainly, the list is not complete, but it is sufficient to describe the address of such a formation. We find it impossible to prepare technical solutions that definitively defeat these dangers, and this leads us to focus more on awareness. And whoever does not make it or does not want to, remain completely out of it. We can be sure that widespread digital hygiene will definitively defeat dangers not destined for a specific target, and these are to a much lesser extent, and generally independent of a digitalized scenario. Put simply, there cannot be efficient digitalization without corresponding digital hygiene. Of course, several companies that have special security needs will need more, but in any case, they too will benefit from digital hygiene; they will only have to add other, more specific, and technical measures.

Another point is related to education. We see two types of education, one for the management, another for the final users. The two paths should not be confused, being the former more interested in governance, policies, and other high-level questions. The latter should be oriented to create the digital hygiene, awareness, etc., being much more operational. And the educators, although the ease of the subjects, should be experienced people, because needing to be able to view the subject even with the eyes of a manager/user, what exactly comes from a long experience.

## 4. CONCLUSIONS

What we have described is not a result, rather it is a meta-result: addressing a new line of research for benefitting at most from digitalization. In Fig. 1 we try to explain how we see the digitalization process, heavily using information technology resources but also deeply entering the several – different – application domains. We believe that this is the task of an information technology expert, assisted by others, including those in the application domain. A computer scientist's view of the scenario is undoubtedly the most complete and correct. (S)he certainly has the competence and sensitivity necessary for this goal, including the important ones of reducing problems, errors, and making all processes more efficient. Avoiding the vendor lock-in is today not only a clear requirement, but first something coming from the experience/errors. And only an expert can assist in that.

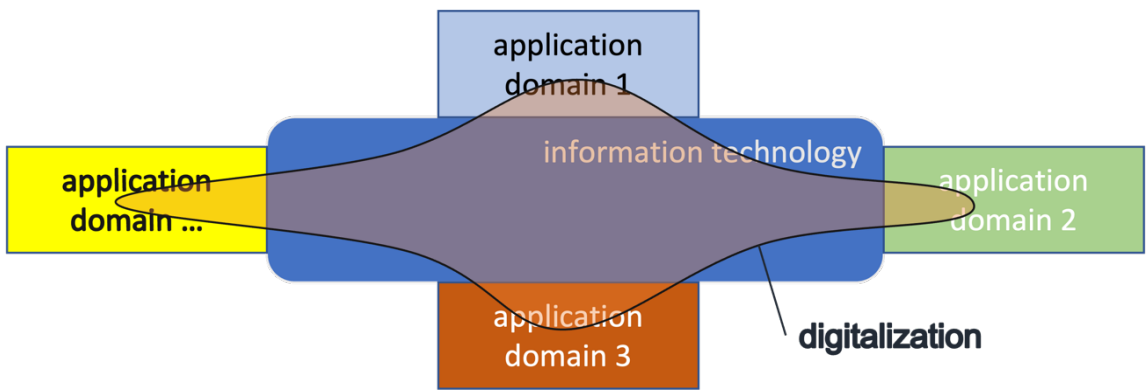

Figure 1. Relationships between information technology, digitalization, and application domains. In blue: information technology; in transparent orange: the digitalization; in light blue, green, red, and yellow: the application domains.

Be careful not to confuse these notes with the traditional computer science research, which will continue to develop results and products; let's think of machine learning, computer vision, robotics, cybersecurity, theoretical computer science, software management models, etc., which will continue to be areas of great interest in computer science. Here we are only proposing a further point on which to discuss and work. Soon we intend to produce precise specifications on tools and procedures to be used to eliminate the errors described and test some existing solutions, perhaps adapting them to our needs.

We don't pretend to include all significant aspects of the question and we have limited ourselves to collecting a handful of obvious requirements, with a view to making the most of the experience of the past in order not to stumble over the same mistakes and to focus on a post-pandemic world that has been able to take advantage of the circumstance to make better use of pre-existing technologies.

## ACKNOWLEDGEMENTS

This work has been partially supported by the IoT-STYLE project RG12117A7CE68848.

**Author**

Short Biography

Fabrizio d'Amore is an associate professor at Sapienza University of Rome, Italy. He teaches Theoretical Computer Science and Cybersecurity. He is responsible for the graduated master program "Information Security and Strategic Information". His teaching/scientific interests include Theoretical Computer Science, Digitalization, Applied Cryptography, Privacy, and Algorithms. He also serves for many public Italian bodies.

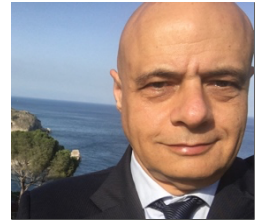